\documentclass[showpacs,showkeys,preprintnumbers,amsmath,amssymb]{revtex4}

\usepackage{graphicx}
\usepackage{dcolumn}
\usepackage{bm}
\begin{document}
\title{Nuclear cross sections in $^{16}\text{O}$ for $\beta$ beam neutrinos at intermediate energies}
\author{S. K. Singh$^{a,b}$, M. Sajjad Athar$^{a}$ and Shakeb Ahmad$^{a}$}
\affiliation{$^{a}$Department of Physics, Aligarh Muslim University, Aligarh- 202002, India.\\
$^{b}$Departmento de Fisica Teorica and IFIC, Centro Mixto Universidad de Valencia CSIC; Institutos de Investigacion de Paterna, Aptd 22085, 46071 Valencia, Spain.}
\begin{abstract}
The nuclear cross sections for charged lepton production induced by $\beta$ beam neutrinos (anti-neutrinos) in $^{16}$O have been presented at intermediate energies corresponding to the Lorentz boost factor $\gamma<250 (150)$. The calculations for quasi-elastic lepton production includes the effect of Pauli blocking, Fermi motion and renormalization of weak transition strengths in the nuclear medium. The calculations for the inelastic lepton production is done in the $\Delta$ dominance model. The renormalization of $\Delta$ properties in a nuclear medium and final state interactions of pions with the final nucleus are taken into account. The results may be useful in performing feasibility studies for the future CERN-FREJUS base line neutrino oscillation experiments.
\end{abstract}
\maketitle
Neutrino experiments done with atmospheric~\cite{fukuda}, accelerator~\cite{ahn}, reactor~\cite{eguchi} and solar~\cite{fukuda1} neutrinos provide evidence for neutrino oscillations. In a three flavor oscillation scenario for the Dirac neutrinos, the three neutrino masses (${m}_i$, i=1,2,3) and mixing angles $\theta_{ij}$(i$\ne$j=1,2,3) and a CP violating phase $\delta$ have to be determined. The present experiments provide limits on $\Delta {m}^2_{12}$, $\theta_{12}$, $\Delta {m}^2_{23}$ and $\theta_{23}$, while the mixing angle $\theta_{13}$ is poorly determined and the $\delta$ phase is still unknown. In addition, the hierarchal structure of $\Delta {m}^2_{ij}$ and the absolute scale of neutrino masses have also to be determined. The high precision neutrino experiments to be performed in the future are expected to improve the present limits on the various parameters of three flavor neutrino oscillation phenomenology. For the purpose of future long base line neutrino oscillation experiments, new sources of neutrino beams like neutrino factories~\cite{geer}, superbeams~\cite{itow} and $\beta$-beams~\cite{zuc} have been proposed. One of these sources, the $\beta$-beams, provide a source of pure single flavor, well collimated and intense neutrino(antineutrino) beams with a well defined energy spectrum obtained from the $\beta$-decay of accelerated radioactive ions boosted by a suitable Lorentz factor $\gamma$. The radioactive ion and the Lorentz boost factor $\gamma$ can be properly chosen to provide the low energy~\cite{volpe3}-\cite{volpe1}, intermediate and high energy~\cite{mez}-~\cite{terranova} neutrino beams according to the needs of a planned experiment.

In the feasibility study of $\beta$-beams, $^{6}$He ions with a ${Q}$ value of 3.5 MeV and $^{18}$Ne ions with a ${Q}$ value of 3.3 MeV are considered to be the most suitable candidates to produce antineutrino and neutrino beams~\cite{autin}. The possibility of accelerating these ions using the existing CERN-SPS, upto its maximum power enabling it to produce beta beams with $\gamma$= 150 (250) for $^{6}$He($^{18}$Ne) ions has been discussed in the literature~\cite{mez1},~\cite{burguet1} which may be used to plan a base line neutrino experiment at L=130 km to the underground Frejus laboratory with the 440 kT water Cerenkov detector~\cite{mez}-~\cite{mez2}. The feasibility of such an experimental setup and its response to $\beta$-beam neutrinos corresponding to various values of the Lorentz boost factor $\gamma$ has been studied by Autin et al.~\cite{autin}. In the range of high $\gamma$, this provides greater sensitivity to the determination of the mixing angle $\theta_{13}$ and the CP violating phase angle $\delta$~\cite{burguet}. In addition, such a facility is also expected to provide the low energy neutrino nuclear cross sections corresponding to very low $\gamma$, which may be useful in calibrating various detectors planned for the observation of supernova neutrinos~\cite{volpe3},~\cite{jacho} and neutrinoless double $\beta$-decay~\cite{volpe1}.

In this letter, we discuss the nuclear response for the $\beta$-beam neutrinos (antineutrinos) of intermediate energy corresponding to the various values of $\gamma$ discussed in the literature. In particular, we study the neutrino nucleus interaction cross sections in $^{16}$O for $\beta$-beam neutrino(antineutrino) energies corresponding to the Lorentz boost factor $\gamma$ in the range of 60$<\gamma<$250 (150). The energy spectrum of $\beta$-beam neutrinos(antineutrinos) from $^{18}$Ne($^{6}$He) ion source in the forward angle($\theta=0^o$) geometry, corresponding to the Lorentz boost factor $\gamma$ is given by~\cite{serreau}:
\begin{eqnarray}
&&\Phi_{lab}(E_\nu,\theta=0)=\frac{\Phi_{cm}(E_\nu\gamma[1-\beta])}{\gamma[1-\beta]}\nonumber\\
&&\Phi_{cm}(E_\nu)=bE^2_\nu E_e p_e F(Z^\prime, E_e)\Theta(E_e-m_e)
\end{eqnarray}
In the above equation $b=ln2/m^5_eft_{1/2}$ and $E_e(=Q-E_\nu)$, $p_e$ are the energy and momentum of the outgoing electron, Q is the Q value of the beta decay of the radioactive ion $A(Z,N)\rightarrow A(Z^\prime,N^\prime) +e^-(e^+) +{{\bar\nu}_e}(\nu_e)$ and $F(Z^\prime, E_e)$ is the Fermi function. In Fig. 1, we show the representative spectra for neutrinos(antineutrinos) corresponding to the Lorentz boost factor $\gamma= $250 (150).
\begin{figure}[h]
\includegraphics{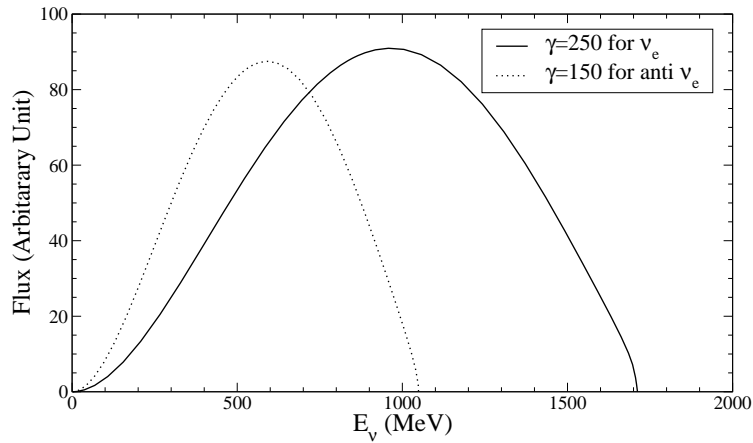}
\caption{Neutrino(solid line) energy spectrum obtained with $^{18}$Ne boosted at $\gamma=250$ and antineutrino(dashed line) energy spectrum obtained with $^{6}$He boosted at $\gamma=150$.}
\end{figure}
In this energy region the dominant contribution to the charged lepton production cross section comes from the quasielastic reactions. However, the high energy neutrinos corresponding to the tail of an energy spectrum, specially for higher $\gamma$ (see Fig.1), can contribute to the inelastic production of charged leptons through the excitation of the $\Delta$-resonance. In addition to the genuine inelastic production of the charged leptons which will be accompanied by the pions, the neutral current induced inelastic production of $\pi^0$ without any charged lepton in the final state can mimic the quasielastic production of charged leptons in which one of the photons from the $\pi^0$ decays is misidentified as a signature of the quasielastic electron production. We, therefore, study the quasielastic and the inelastic production of charged leptons induced by the charge current. We also study the neutral current induced production of $\pi^0$ which gives major contribution to the background of the electron production in the quasielastic reactions induced by neutrinos and antineutrinos.

The cross section for quasielastic charged lepton production for the process $\nu_e$ + $^{16}$O $\rightarrow e^- +$ $^{16}$F$^\star$ is calculated in a local density approximation using the standard model Lagrangian for the weak interaction using Budd, Bodek and Arrington (BBA03)~\cite{budd} weak nucleon axial vector and vector form
factors with ${M}_{A}$=1.05 GeV and ${M}_{V}$=0.84 GeV. The Fermi motion and Pauli blocking effects in nuclei are included through the imaginary part of the Lindhard function for particle hole excitations in the nuclear medium. The renormalization of weak transition strengths, which are quite substantial in the spin-isospin channel, are calculated in the random phase approximation(RPA) through the interaction of p-h excitations as they propagate in the nuclear medium using a nucleon-nucleon potential described by pion and rho exchanges. The effect of  Coulomb distortion of the electron in the field of the final nucleus is also taken into account by using a local version of the modified effective momentum approximation~\cite{engel}-\cite{singh1}. The details of the formalism and the relevant expressions for the cross section are given in refs.~\cite{singh1}-\cite{nieves}. 

The cross section for inelastic charged lepton production for the process $\nu_e(\bar{\nu}_e)$ + $^{16}$O $\rightarrow e^-(e^+) + \pi^{\alpha} + X$, where $\alpha$ is the charge state of the pion, is calculated in the $\Delta$ dominance model using a local density approximation. The sequential production of pions through the excitation of the $\Delta$ resonance and its subsequent decay in pions through the $\Delta\rightarrow{N}\pi$ process is considered. The $\Delta$ resonance is described by a Rarita Schwinger field and the matrix element for the $\Delta$ excitation is written  using the weak N$\Delta$ transition form factors which are determined from the analysis of the data available on the photo-, electro- and neutrino- excitation of  the $\Delta$ resonance. The use of CVC along with the experimental data on electromagnetic excitation of the $\Delta$ is used for determining the vector form factors while the hypothesis of PCAC along with the experimental data on neutrino excitation of $\Delta$ from $\nu_\mu-{d}$ reactions have been used to determine the axial vector form factors. The matrix elements and the form factors have been discussed in refs.~\cite{singh3}-\cite{lalakulich}. The effect of a nuclear medium on the width and mass of the $\Delta$ is included in a model where the self energy of the $\Delta$ in nuclear medium is calculated in a local density approximation~\cite{singh3},~\cite{oset}. The final state interaction of pions with the final nucleus is described by using energy dependent pion absorption probability provided by Vicente Vacas~\cite{vicente1}-~\cite{vicente}. This formalism is also applied to calculate the neutral current induced $\pi^0$ production process i.e. $\nu_e$ + $^{16}$O $\rightarrow \nu_e + \pi^{0} + X$ in order to study the major source of background to the quasielastic charged lepton events~\cite{mez2}.

The numerical calculations for the total cross section $\sigma(E_\nu)$ in $^{16}$O have been made using the 3-parameter Fermi density $\rho(r)$ given by\cite{vries}:
\[\rho(r)=\rho_0\left(1+w\frac{r^2}{c^2}\right)/\left(1+exp\left(\frac{r-c}{z}\right)\right)\]
with c= 2.608 fm, z= 0.513 fm and w= -0.051 and the results have been presented in Fig.2 and Fig.3 for the neutrino and the antineutrino reactions. We see from Fig.2 that for the neutrino reactions, the charged lepton production is dominated by the quasielastic production and the inelastic charged lepton production becomes comparable only around $E_\nu\sim 1.5$~GeV. The neutral current inelastic production of $\pi^0$ is small and is about $12-15\%$ of the quasielastic charged lepton production in the energy range of 0.8 GeV$<E_\nu<$1.0 GeV. Therefore, the background to the quasielastic lepton events due to the neutral current $\pi^0$ production is expected to be important only at high $\gamma$ (for example $\gamma$=250) where it could be around $15\%$ corresponding to the average neutrino energies  $E_\nu\sim 1.0$~GeV. Qualitatively, similar results are obtained for the antineutrino reactions and are shown in Fig.3.
\begin{figure}[h]
\includegraphics{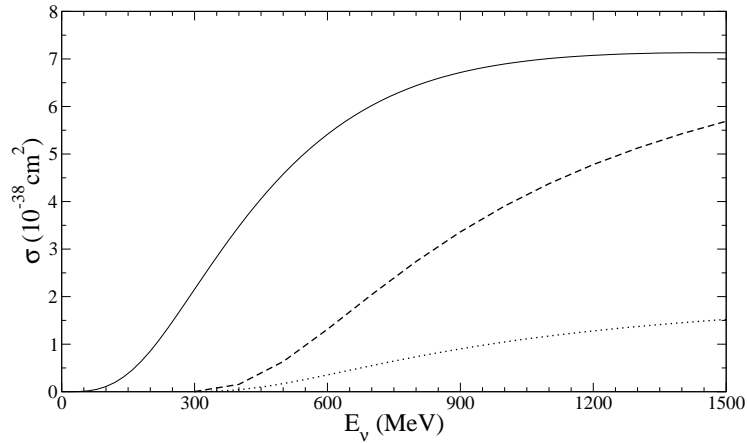}
\caption{Total cross section $\sigma(E_{\nu_e})$ vs $E_{\nu_e}$ for the neutrino reaction in $^{16}$O for the quasielastic(solid line), inelastic charged lepton production processes(dashed line), and inelastic neutral current production of $\pi^0$(dotted line).}
\end{figure}
\begin{figure}[h]
\includegraphics{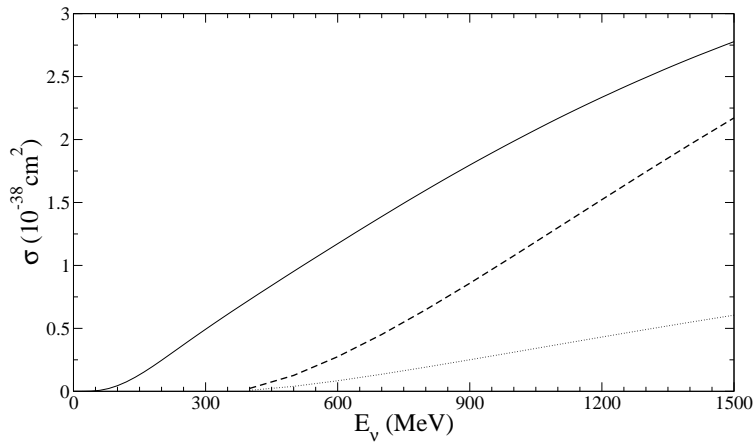}
\caption{Total cross section $\sigma(E_{{\bar\nu}_e})$ vs $E_{{\bar\nu}_e}$ for the antineutrino reaction in $^{16}$O for the quasielastic(solid line), inelastic charged lepton production processes (dashed line), and inelastic neutral current production of $\pi^0$ (dotted line).}\end{figure}

We would like to emphasize that nuclear medium effects play an important role in reducing the cross sections specially for the quasielastic charged lepton production in the low energy region. For example, we find that with the Pauli blocking and the Fermi motion of the nucleons in the nucleus the cross sections reduce from the free case by $30~\%$ at $E_{\nu_e}=200$~MeV, $15\%$ at $E_{\nu_e}=400$~MeV, $10\%$ at $E_{\nu_e}=750$~MeV and around $9\%$ at $E_{\nu_e}=1.0$~GeV. When the RPA correlation in the nuclear medium is also taken into account there is a total reduction of $60\%$ at $E_{\nu_e}=200$~MeV, $40\%$ at $E_{\nu_e}=400$~MeV, $26\%$ at $E_{\nu_e}=750$~MeV and around $23\%$ at $E_{\nu_e}=1.0$~GeV. In the case of inelastic charged lepton production, we find that when nuclear medium modification effects on the $\Delta$ properties are taken into account,the cross section reduces by around $15\%$ for energies $E_{\nu_e}=0.5-1.0$~GeV as compared to the cross sections calculated without the medium modification effects. When the final state interaction of pions with the residual nucleus is taken into account there is a further reduction in the cross section which leads to a total reduction of around $40\%$ for the neutrino energies $E_{\nu_e}=0.5-1.0$~GeV. Similar results are also obtained for the neutral current $\pi^0$ production. 

The effect of nuclear medium on neutrino induced quasielastic production of leptons from $^{16}$O in the intermediate energy region of present interest has been studied by many other authors~\cite{nieves} and ~\cite{smith}-\cite{mai} and our results  for total cross sections presented in Figs. 2 and 3 are in qualitative agreement with the results of Valverde et al.~\cite{nieves}, Gaisser and O'Connell~\cite{gas}, and Marteau~\cite{marteau} but are smaller than the results of Maieron et al.~\cite{mai}. In the case of inelastic neutrino lepton production induced by charged currents in this energy region, the effect of nuclear medium and final state interactions has been studied earlier by some authors~\cite{singh3},~\cite{marteau},~\cite {kim}, and~\cite{paschos}. The results presented here for nuclear medium effects in the total cross sections corresponding to the charged current inelastic lepton production and the neutral current pion production are consistent with our earlier results~\cite{singh3} and the ones of Marteau~\cite{marteau} and Kim, Schramm and Horowitz~\cite{kim}. We can not compare our results to those of Paschos and collaborators~\cite{paschos} because they present results for the  momentum and angular distributions and do not report results on total cross sections.  

We have also calculated the total cross sections for the coherent production of charge and neutral current pions in this energy region and find that the cross sections are quite small as compared to the incoherent production cross sections presented here. This has been discussed in refs.~\cite{singh6}-~\cite{paschos1} where the theoretical results for the coherent production of pions have been found to be in reasonable agreement with the preliminary experimental results reported by the  K2K~\cite{k2k} and the MiniBooNE~\cite{boone} collaborations. Therefore, the coherent pion productions are not expected to give any significant contribution to the number of charged lepton events in this energy region.

In order to estimate the relative contribution of the quasielastic and the inelastic production of charged leptons and also the background to the quasielastic events due to the neutral current induced neutral pion production at a far detector in a base line experiment, we have calculated the flux averaged cross section $\langle\sigma\rangle$ defined as
\begin{equation}
\langle\sigma\rangle=\frac{\int^{\infty}_0 dE_\nu\Phi_{lab}(E_\nu,\theta=0)\sigma(E_\nu)}{\int^{\infty}_0 dE_\nu\Phi_{lab}(E_\nu,\theta=0)}
\end{equation}
for the neutrino and the antineutrino energies. This is relevant for the future CERN-FREJUS base line experiments which can be done with the present CERN-SPS and have been discussed in the literature~\cite{mez1},~\cite{burguet}. The forward angle approximation for the neutrino flux used in equation(2) to calculate the total cross section is quite good for a far detector specially for higher values of the Lorentz factor $\gamma$. Quantitatively, we find that the contribution to the total cross section from non-zero $\theta$ flux i.e. $\Phi_{lab}(E_\nu,\theta\ne 0)$ is about $5\%$ for $\gamma=60$ and reduces to less than $1\%$ for $\gamma=250$. In Tables 1 and 2, we show the results of the flux averaged cross section $\langle\sigma\rangle$ for neutrino and antineutrino reactions for various values of the Lorentz boost factor $\gamma$ where we can see the relative contributions of the cross sections for quasielastic and inelastic production of leptons along with the cross sections for neutral current induced production of neutral pions which is the major source of background to the quasielastic events at intermediate energies.

To summarize, we have presented in this letter the numerical results for the charged current lepton production induced by $\beta$-beam neutrinos (antineutrinos) in $^{16}$O, calculated in a local density approximation which takes into account the nuclear medium effects. The calculations have been done for the quasielastic production of charged leptons using RPA and the inelastic production of charged leptons  using $\Delta$ dominance model. The renormalization of the $\Delta$ properties in a nuclear medium is included through the self energy of the $\Delta$ in the nuclear medium calculated in a local density approximation. The neutral current induced neutral pion production, which constitutes the major background to the quasielastic charged lepton events, is also calculated in this model.

We would like to thank M. J. Vicente Vacas for providing the pion absorption probabilities and H. Arenhoevel for reading the manuscript.The work is supported by the Department of Science and Technology, Government of India under the grant DST Project No. SP/S2K-07/2000. One of the authors (S. Ahmad) would like to thank CSIR for the financial support.
\begin{table}[h]
\caption{Cross sections ${\langle\sigma\rangle}_{\nu_e}$ averaged over the $\beta$ beam neutrino spectrum for various Lorentz boost factor $\gamma$(column I) and corresponding average energies of neutrinos (column II). Columns III and IV give the total cross sections for the quasielastic and the inelastic charged lepton production process and column V gives the the total cross section for the inelastic neutral current production of $\pi^0$.}
\begin{center}
\begin{tabular}{ccccc}\\ \hline
&&${\langle\sigma\rangle}_{\nu_e}$ in $10^{-40}{cm}^2$&&\\ \hline
 $\gamma$&$\langle E_{\nu_e}\rangle$&${\langle\sigma\rangle}_{qe}^{cc}$&${\langle\sigma\rangle}_{inel}^{cc}$&${\langle\sigma\rangle}_{inel}^{nc}$\\
\hline
60& 226  & 131 & 0.87   &0.24 \\
75& 282  & 199 & 6.3   &1.7  \\
100& 376 & 307 & 33   &9 \\
150& 564 & 468 & 131   &35.3  \\
200& 752 & 563 & 238   &64    \\
250& 940 & 617 & 331   &88.7    \\ \hline
\end{tabular}
\end{center}
\end{table}
\begin{table}[h]
\caption{Cross sections ${\langle\sigma\rangle}_{{\bar\nu}_e}$ averaged over the $\beta$ beam antineutrino spectrum for various Lorentz boost factor $\gamma$(column I) and corresponding average energies of antineutrinos (column II). Columns III and IV are give the total cross sections for the quasielastic and the inelastic charged lepton production process, while column V gives the total cross section for the inelastic neutral current production of $\pi^0$.}
\begin{center}
\begin{tabular}{ccccc}\\ \hline
&&${\langle\sigma\rangle}_{{\bar\nu}_e}$ in $10^{-40}{cm}^2$&&\\ \hline
 $\gamma$&$\langle E_{{\bar\nu}_e}\rangle$&${\langle\sigma\rangle}_{qe}^{cc}$&${\langle\sigma\rangle}_{inel}^{cc}$&${\langle\sigma\rangle}_{inel}^{nc}$\\
\hline
60& 232  & 33.2 & 0.0855 &0.028 \\
75& 290  & 46.6 & 1.2   &0.38  \\
100& 387 & 69 & 7.2   &2.2  \\
150& 580 & 111 & 32     &9.5  \\ \hline
\end{tabular}
\end{center}
\end{table}

\end{document}